\documentclass[galaxies,review,accept,pdftex,moreauthors]{Definitions/mdpi} 

\newcommand\aap{A\&A}                
\newcommand\aapr{A\&ARv}             
\newcommand\aj{AJ}                   
\newcommand\apj{ApJ}                 
\newcommand\apjl{ApJ}                
\newcommand\apjs{ApJS}               
\newcommand\araa{ARA\&A}             
\newcommand\mnras{MNRAS}             
\newcommand\nar{New~Astron.~Rev.}    
\newcommand\nat{Nature}              
\newcommand\rmxaa{Rev. Mex. Astron. Astrofis.} 
\newcommand\ssr{Space Sci. Rev.}     

\firstpage{1} 
\pubvolume{1}
\issuenum{1}
\articlenumber{0}
\pubyear{2024}
\copyrightyear{2024}
\datereceived{28 February 2024} 
\dateaccepted{3 April 2024} 
\datepublished{5 April 2024} 
\hreflink{https://doi.org/} 

\Title{Observational Tests of AGN Feedback: An Overview of Approaches and Interpretation}

\TitleCitation{Observational Tests of
Active Galactic Nuclei Feedback: An
Overview of Approaches and
Interpretation}

\Author{Chris M. Harrison $^{1}$\orcidA{} and Cristina Ramos Almeida$^{2,3}$\orcidB{}}

\AuthorNames{Chris M. Harrison \& Cristina Ramos Almeida}

\AuthorCitation{Harrison, C.M.; Ramos Almeida, C.}

\address{%
$^{1}$ \quad School of Mathematics, Statistics and Physics, Newcastle University, NE1 7RU, UK\\
$^{2}$ \quad Instituto de Astrofísica de Canarias, Calle Vía Láctea, s/n, 38205 La Laguna, Tenerife, Spain\\
$^{3}$ \quad Departamento de Astrofísica, Universidad de La Laguna, 38206 La Laguna, Tenerife, Spain
}

\corres{Correspondence: christopher.harrison@newcastle.ac.uk; cra@iac.es}

\abstract{Growing supermassive black holes (Active Galactic Nuclei; AGN) release energy with the potential to alter their host galaxies and larger-scale environment; a process named ``AGN feedback''. Feedback is a required component of galaxy formation models and simulations to explain observed properties of galaxy populations. We provide a broad overview of observational approaches that are designed to establish the physical processes that couple AGN energy to the multi-phase gas, or to find evidence that AGN impact upon galaxy evolution. The orders-of-magnitude range in spatial, temporal, and temperature scales, requires a diverse set of observational studies. For example, studying individual targets in detail sheds light on coupling mechanisms; however, evidence for long-term impact of AGN is better established within galaxy populations that are not necessarily currently active. We emphasise how modern surveys have revealed the importance of radio emission for identifying, and characterising, feedback mechanisms. At the achieved sensitivities, the detected radio emission can trace a range of processes, including shocked interstellar medium caused by AGN outflows (driven by various mechanisms including radiation pressure, accretion disc winds, and jets). We also describe how interpreting observations in the context of theoretical work can be challenging, in part, due to some of the adopted terminology.}


\begin{document}

\section{Introduction}

The presence of supermassive black holes (SMBHs) at the centre of galaxies, including our own Milky Way, is confirmed by measurements of the motions of stars and gas in galaxy centres, and by the gravitational lensing signatures within multi-wavelength images (e.g., \cite{ghez08,kormendy13,eht19,gravity20,nightingale23}). These black holes (BHs) grow through periods of gas accretion, which are known as Active Galactic Nuclei (AGN), and are identified using observations across the electromagnetic spectrum \cite{soltan82,marconi04,alexander12}. The energy released by AGN was discussed as potentially important for influencing galaxy evolution as early as the 1980s; however, it is over the last three decades that there has been prolific research into this so-called ``AGN feedback'' process (see discussion in \cite{harrison18}).

The explosion of interest in AGN feedback was largely triggered by: (1) the observed correlation between black hole masses and stellar bulge properties \cite{magorrian98,gebhardt00,ferrarese00}, which early analytical models sought to explain with the effect of AGN driving gaseous outflows into their hosts \cite{silk98,king03,king05}; (2) the lower-than-expected rate of gas cooling identified around the most massive galaxies, which implies a heating source that is attributed to AGN \cite{binney95,ciotti97,peterson03} and; (3) semi-analytical models and hydrodynamic simulations, which invoked AGN to regulate black hole growth and to reduce the efficiency of star formation (SF) at the highest stellar masses, as a required process to successfully reproduce the observed local stellar mass functions, and colour bi-modality, of galaxies \cite{benson03,springel05,croton06,bower06,somerville08}.

Whilst energetic feedback from stellar winds and supernovae are typically considered sufficient for low mass galaxies (with stellar masses $\lesssim$10$^{10}$ solar masses), AGN feedback remains a crucial component of galaxy evolution theory to explain the massive end of the galaxy population. This is demonstrated by its ubiquitous role within the current generation of cosmological simulations (e.g., \cite{schaye15,khandai15,dubois16,naab17,mccarthy17,nelson18,dave19}). Across these simulations, AGN remain the only known solution to regulate the cooling in dense galaxy environments, and to sufficiently control the efficiency of star formation for massive galaxies. Furthermore, within modern models and simulations, AGN are considered important for establishing other observed properties of galaxies and their environments. This includes galaxy structures, mass profiles and dynamics, and many of the properties of the gas observed in and around galaxies (e.g., \cite{mccarthy17,choi18,davies19,periani19,vandervlugt19,wright20,cochrane23,obreja24}).

Observations are required to test any theoretical ideas and implementations of AGN feedback within simulations, whilst carefully keeping in mind the difference between true predictions and observations that were used to calibrate the models (e.g., \cite{schaye15}; see Sections~\ref{sec:energetics} and \ref{sec:longterm}). Furthermore, observations should be used to constrain the physical mechanisms by which the energy released by AGN is able to couple to the gas. Indeed, there is currently no theoretical consensus on this, with an ongoing discussion on the relative importance of different mechanisms including: radiation pressure on dusty gas; accretion disc winds; radio jets; and cosmic rays (e.g., \cite{zakamska16,ishibashi18,rakshit18,jarvis19,costa20,bruggen20,venturi21,mandal21,meena23,calistrorivera24}). Consequently, the literature is full of observational studies exploring these topics. They range from multi-wavelength, highly detailed, observations of individual targets through to statistical tests on large galaxy samples. Due to the diversity of these studies, it can be challenging to interpret the results within the context of one another, and within the context of the theoretical ideas around AGN feedback. We aim to address this challenge in this article by providing a broad overview of the different observational approaches to test AGN feedback, and discuss how they can (or can not) be related to one another and to theoretical predictions. 

Observational work on AGN feedback has been reviewed multiple times over the last decade, including by \cite{alexander12,fabian12,mcnamara12,kormendy13,heckman14,harrison17,morganti17,king15,harrison18,hardcastle20,veilleux20,bourne23,krause23}. These reviews typically focus in depth on one aspect of the feedback processes (e.g., the role of jets or the properties of the multi-phase gaseous outflows), and/or they are dedicated to one particular sub-set of the AGN population (e.g., radio-detected AGN, those in the densest galactic environments, or the most luminous AGN). This article does not aim to supersede these previous focused reviews. Instead, we aim to complement them by providing a more holistic overview of the breadth of observational approaches taken to search for, and to characterise, AGN feedback. The goal is to place the different observational approaches within the context of one another, and to comment upon how they collectively can be used to test, or refine, different theoretical ideas of feedback.


\section{A multi-faceted approach to a complex problem}\label{sec:overview}

AGN feedback is a problem of many scales. Both black hole growth (AGN) and stellar growth (star formation) are fed by gas. Therefore, we are interested in understanding a balance of feeding from, and feedback upon, a gaseous fuel supply. As discussed below, the relationship between these processes is a multi-scale problem over many orders of magnitude in spatial, temperature, and temporal scales. This range of scales is represented in the schematic diagram shown in Figure~\ref{fig:TDscales}.

\subsection{Multiple spatial scales, temperatures, gas phases, and timescales}\label{sec:spatialscales}

To assess the relationship between AGN and galaxy formation, we need to understand physical processes that can be occurring over greater than six orders of magnitude in spatial scales. The spatial scales of accretion flows onto SMBHs are sub-parsec. Indeed, the physics that determine both the final accretion on to black holes and the corresponding energy output can take place on extremely small scales, down to the innermost stable circular orbit. These accretion flows are fed from the surrounding {\em circumnuclear region} (CNR), which is considered to be on scales of $\sim$1--10\,pc, and is also responsible for significant obscuration of the central AGN (i.e., the torus  \cite{ramosalmeida17}). Therefore, the interaction between the AGN and the material in the CNR can determine both the rate of black hole growth and the ability to observe AGN emission. Galaxies are typically considered to have their {\em interstellar medium} (ISM) located on scales from a few parsecs up to tens of kiloparsecs, and star formation primarily occurs within this ISM \cite{tacconi20}. If AGN impact directly on the ISM (e.g., by entraining/expelling it, or by heating/exciting/ionising it; see Section~\ref{sec:gasimpact}), they have the potential to impact directly upon the efficiency of star formation within the host galaxy (see Section~\ref{sec:localimpact}). 

\vspace{-0.5cm}

\begin{figure}[H]
\centering
\includegraphics[width=13cm]{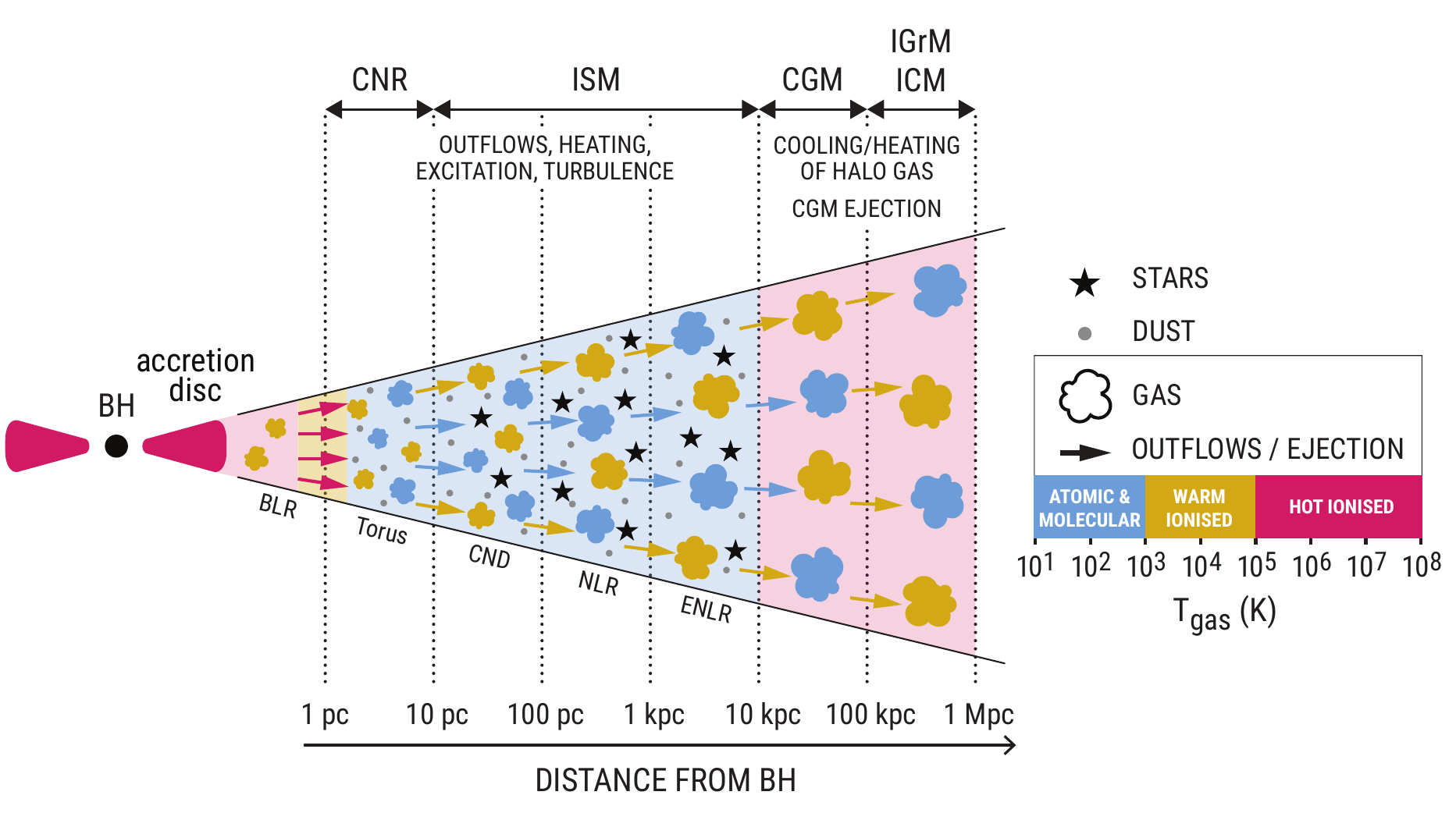}
\caption{A schematic diagram to highlight the orders-of-magnitude range in temperature (colour-coding of gas clouds, intra-clouds medium and outflows) and spatial scales of AGN feedback. Gas in a range of phases, with a huge range in temperatures, can be affected by the energy released by AGN. This can occur from nuclear scales through to galaxy cluster scales. This large range in spatial scales naturally corresponds to an equivalently large range in temporal scales. See discussion in Section~\ref{sec:overview}.
\label{fig:TDscales}}
\end{figure}

The gas surrounding galaxies, outside of the their discs or ISM, but within their virial radii, is known as the {\em circumgalactic region} (CGM; \cite{tumlinson17}). Broadly speaking, this gas extends from $\sim$10\,kpc up to $\sim$100\,kpc. The CGM is a reservoir that can fuel a galaxy's ISM. It can be enriched by material that has been expelled from the host galaxy (e.g., by AGN outflows; Section~\ref{sec:gasimpact}). If an AGN can expel gas from the CGM, this could ultimately prevent or regulate future star formation in the host (e.g., \cite{davies20}). The medium between galaxies that are located within groups or clusters can extend up to hundreds of kiloparsecs. This is called the {\em intragroup medium} (IGrM) or {\em intracluster medium} (ICM), respectively. AGN can act as a source of heating on this gas, and the ability for this gas to cool and flow onto galaxies plays a role in establishing the rate of star formation (see Section~\ref{sec:longterm}).

As summarised in Figure~\ref{fig:TDscales}, the CNR, ISM, CGM, and IGrM/ICM can all contain gas at a range of temperatures, and in a range of phases (i.e., molecular, atomic, or ionised; see e.g., \cite{ramosalmeida17,tacconi20,ganguly23}). For example, on spatial scales ranging from tens to hundreds of parsecs, the cold and dense ISM (i.e., the molecular gas \cite{garciaburillo21}) is generally referred to as the circumnuclear disc (CND). The ionised gas on similar scales and up to $\sim$1 kpc is known as the narrow-line region (NLR), and up to $\sim$10 kpc as the extended-NLR (ENLR; see Figure \ref{fig:TDscales}). It is the coldest and densest gas that is directly associated with star formation, relating to temperatures of 10s to 100s of Kelvin. Therefore, observations that trace atomic and molecular gas are usually considered necessary for measuring the properties of the available fuel supply for star formation (see Section~\ref{sec:localimpact}).  However, much of this gas will have cooled from hotter phases. Furthermore, AGN are expected to heat and expel considerable gas in hot/ionised phases (e.g., \cite{mukherjee18,costa20,bourne23}), which may also cool or help trigger the formation of colder phases (e.g., \cite{richings18}). Therefore, it is necessary to consider a wide range of gas phases. The corresponding temperature range may span $>$6 orders of magnitude. This creates a huge observational challenge. It has only recently become feasible to obtain multi-phase measurements for representative galaxies samples, yet the properties of some gas phases still remain largely unconstrained (see Section~\ref{sec:gasimpact}). 

Due to the orders-of-magnitude range of {\em spatial} scales expected for the AGN feedback process (see Figure~\ref{fig:TDscales}), we naturally have to consider a huge range of {\em time} scales. Furthermore, we observe signatures of historic (not currently active) AGN events and variability in AGN luminosity (e.g., \cite{peterson01,lintott09,mostert23}). Such observations confirm that black hole growth, and the resulting energy output, is a variable and transient process. Indeed, it is better to consider AGN as {\em events} in the life-cycle of a galaxy and not to consider them as {\em objects} that persist in time. 

The relative amount of time that a black hole is accreting (active), compared to not accreting (inactive), is sometimes referred to as the `duty cycle'; although its exact definition varies in the literature. There are potentially multiple relevant time scales (or `duty cycles') for the AGN feedback process, which may also depend on the `accretion mode' (see Section~\ref{sec:demographics}). These can range from variability on timescales of days--months for the observed high energy AGN emission \cite{peterson01}, a proposed typical high-accretion rate period of growth lasting $\sim$10$^{5}$\,years \cite{kingNixon15,schawinski15}, through to the idea that AGN could nearly always be active at low levels for the most massive galaxies \cite{sabater19}. An important example study in the context of understanding AGN feedback, Hickox et~al. \cite{hickox14}, use the simulations of Novak et~al. \cite{novak11} to demonstrate that black holes may switch between active (i.e., luminous AGN) and non active states, on timescales of $\lesssim$1\,Myr. This is far shorter than the typical timescales for star formation of $\gtrsim$100\,Myr. This severely complicates how we can connect a specific observed AGN event to any impact that it may have on the  star formation in its host galaxy (see further discussion in Sections~\ref{sec:localimpact} and \ref{sec:longterm}).

\subsection{The multi-faceted approach to observational work on AGN feedback}\label{sec:multi-faceted}

\begin{figure}[H]
\centering
{\par\includegraphics[width=5cm]{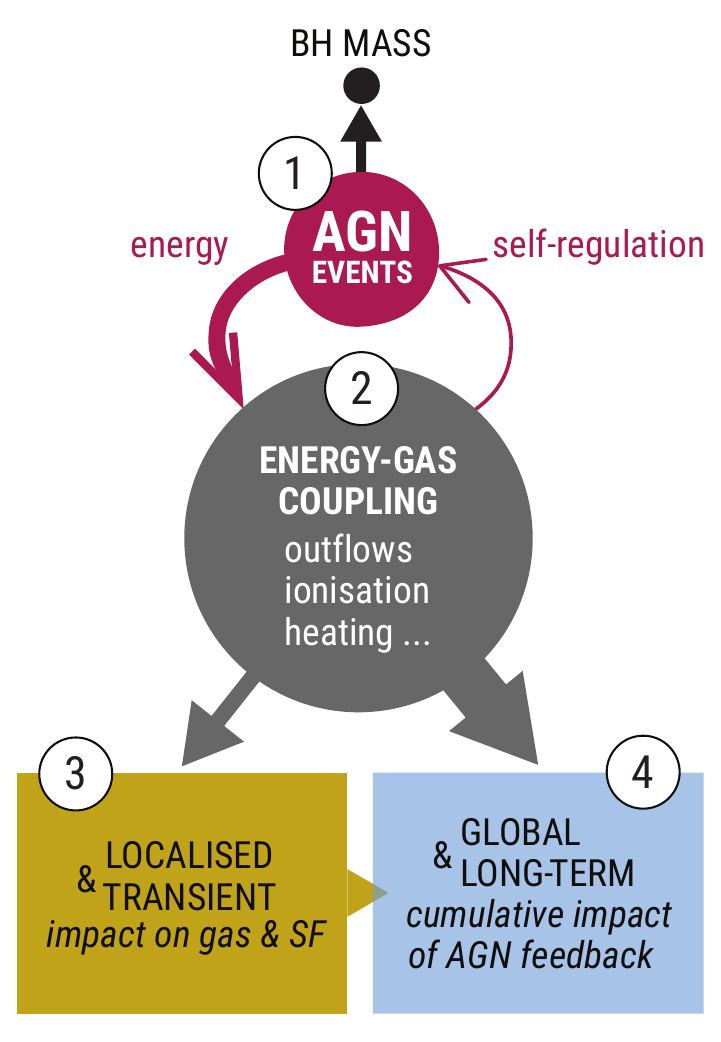}
\includegraphics[width=8.6cm]{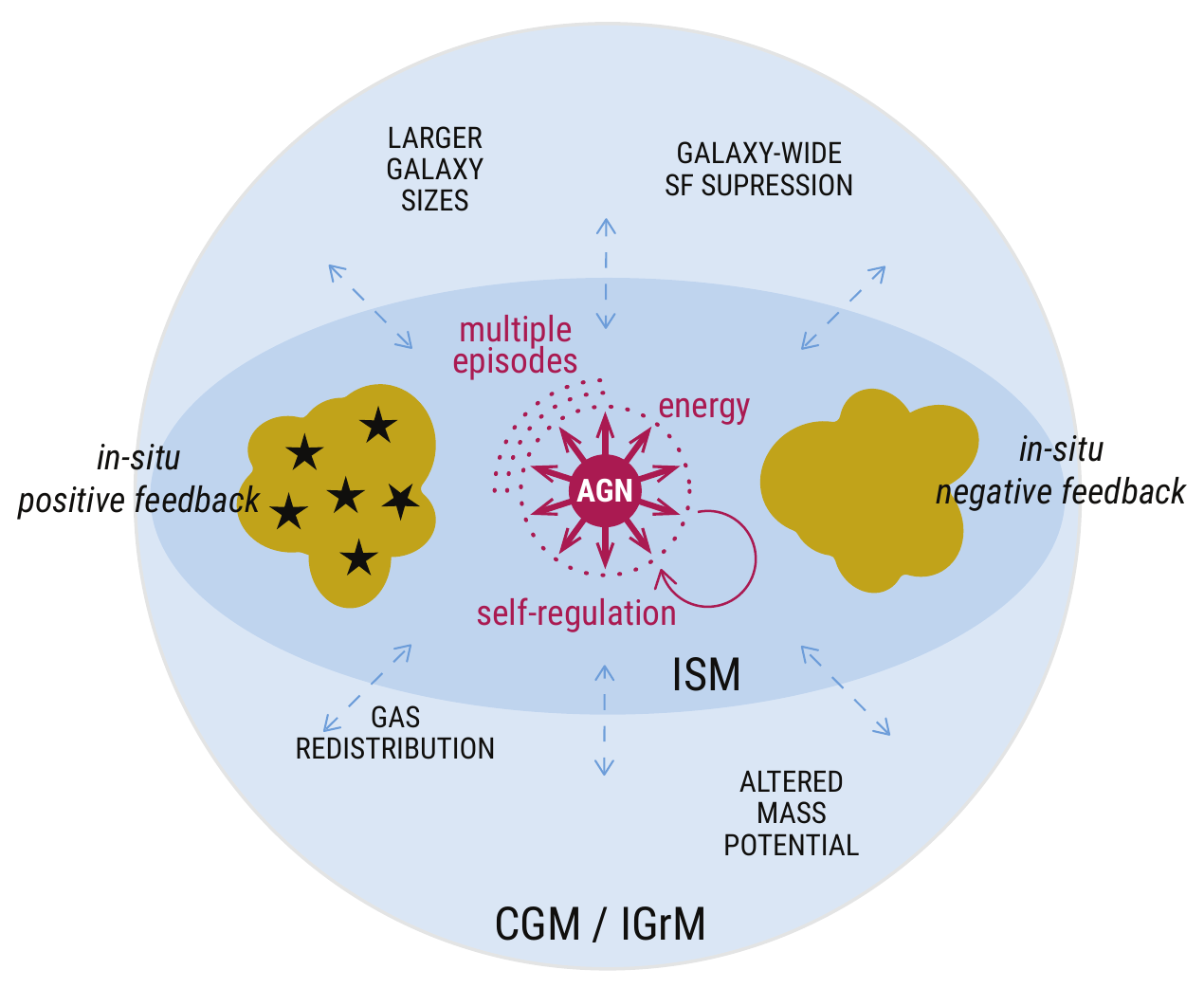}}
\caption{A schematic overview of the different regimes needed to tackle the overall AGN feedback problem with observations (see Section~\ref{sec:multi-faceted}). This is presented in the form of a flow chart (left) and a schematic representation of an AGN-host galaxy embedded in a halo (right). Different colors and numbers within the flow chart correspond to different regions and processes labelled on the right panel and with the same colors. \label{fig:overview}}
\end{figure}

The challenges of scales outlined above, and in Figure~\ref{fig:TDscales}, necessitate a multi-faceted approach to observationally test ideas around AGN feedback. Figure~\ref{fig:overview} provides a schematic overview of the primary regimes involved in addressing this AGN feedback problem. 

Firstly, it is crucial to establish the demographics of AGN across different galaxy populations to assess: (a) the energy budget and duty cycles of AGN events; and (b) the prevalence of different mechanisms of energy injection into the host galaxies (e.g., jets, winds and radiation). This is labelled with a ``1'' in Figure~\ref{fig:overview}, and is discussed further in Section~\ref{sec:demographics}. The energy released by AGN events can have an impact upon the kinematics, distribution, and excitation/ionisation state of the multi-phase gas (labelled with ``2'' in Figure~\ref{fig:overview}). The properties and level of these effects will depend on how efficiently the energy can couple to the gas on various spatial scales. This is important to establish with observations, and is discussed in Section~\ref{sec:gasimpact}. Energy injection into the ISM may have an in-situ, localised affect, on the ability for gas to form stars. This may have a negative or a positive impact on the star formation efficiency (labelled ``3'' in Figure~\ref{fig:overview}) and it is important to understand the details of this process using high angular resolution observations (see Section~\ref{sec:localimpact}). Finally, the cumulative impact of multiple AGN events could have many long-term effects on the global properties of the host galaxies and their larger-scale environment (labelled ``4'' in Figure~\ref{fig:overview}). Assessing the ``imprint'' of AGN feedback on these properties is an important observational test for refining different AGN feedback models invoked in cosmological simulations, and this is discussed further in Section~\ref{sec:longterm}.

\section{AGN accretion modes and methods of energy injection into the host}\label{sec:demographics}

To understand the physical process of AGN feedback, requires an understanding of the prevalence and power of different mechanisms of energy output during AGN events (see region ``1'' in Figure~\ref{fig:overview}; discussed in this section) and constraining how efficiently this energy can couple to the gas (discussed in Section~\ref{sec:gasimpact}). It is beyond the scope of this article to review AGN demographics, but we note the importance of establishing a complete census of AGN events across different galaxy populations and cosmic epochs (see instead, e.g., \cite{alexander12,hickox18,aird18,hardcastle20}). Here we present a broad overview of the mechanisms by which AGN may inject energy into their host galaxies (Section~\ref{sec:energy}), reflect on what can be learnt on this topic from modern-day radio surveys (Section~\ref{sec:radio}) and briefly comment on these in the context of the theoretical `modes' of AGN feedback implemented in some simulations (Section~\ref{sec:energy_terminology}). 

\begin{figure}[H]
\centering
\includegraphics[width=14cm]{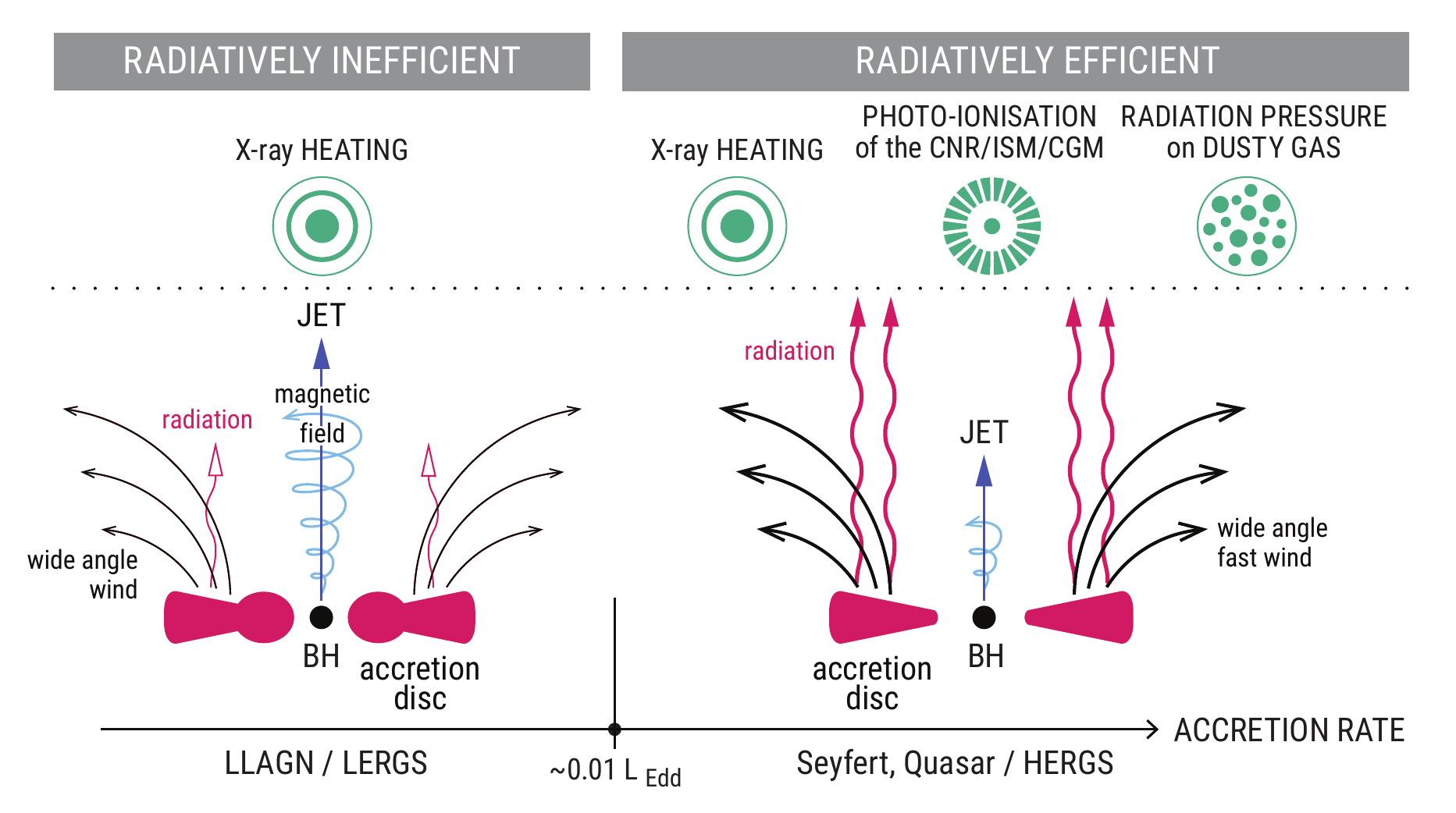}
\caption{A schematic overview of different mechanisms by which AGN can inject energy into the gas in, and around, their host galaxies. This is shown for two regimes of accretion: `radiatively inefficient' for lower accretion rates (left) and `radiatively efficient' for higher accretion rates (right). Multiple mechanisms co-exist in both regimes. This is described in Section~\ref{sec:energy}. \label{fig:powersources}}
\end{figure}

\subsection{AGN accretion modes and mechanisms of energy output}\label{sec:energy}

We will consider two main classes of AGN events: those which are ``radiatively efficient'', where the accreting mass is efficiently converted to radiation, and those which are ``radiatively inefficient'', where the converse is true\footnote{We ignore the possibility of three regimes, with the addition of a distinct extremely highly accretion rate state (`super Eddington'), which may be particularly relevant for early black hole growth, including the origin and accretion rates of massive black holes in the early Universe that have been observed with {\em JWST} \cite{rennehan23,lupi23,maiolino24}.}. Although a simplification, and a matter of ongoing research, we can broadly relate this dichotomy to the accretion rates and the resulting properties of the accretion flow. Sources with the highest accretion rates, i.e., those with luminosities $\gtrsim1$\% of their Eddington luminosity are expected to have a radiatively efficient accretion disc that is geometrically thin and optically thick. In contrast, black holes with lower accretion rates are expected to have radiatively inefficient flows that are geometrically thick and optically thin \cite{shakura73,esin97,narayan94}. The spin of the black holes may also be crucial in determining the mechanical power output available through jets \cite{blandford77,sikora07}. Potential methods of energy injection from these two AGN classes are summarised in Figure~\ref{fig:powersources}.

For radiatively inefficient AGN events (left side of Figure~\ref{fig:powersources}; sometimes called Low Luminosity AGN; LLAGN), the mechanical energy output greatly exceeds the radiative output. This is likely to be dominated by collimated jets of charged particles, although there can be a non-negligible contribution of magnetically- or thermally-driven isotropic (wide-angle) winds from the accretion flow and from Compton (X-ray) heating (e.g., \cite{blandford82,begelman83,yuan14,gan14,hardcastle20,almeida23}). For radiatively efficient AGN\footnote{The variety of multi-wavelength methods used to identify radiatively efficient AGN results in a menagerie of AGN classifications and terminology used in the literature \cite{padovani16}. Here we consider two broad classes: quasars and Seyferts. Quasars have high bolometric luminosities ($L_{\rm AGN}\gtrsim$10$^{45}$\,erg\,s$^{-1}$) and lower luminosity AGN are considered to be Seyferts. Both classes can be type-1, where direct accretion emission, and the broad line region (BLR), is observed, or type-2 where no accretion emission or BLR is detected.} (right side of Figure~\ref{fig:powersources}), the energy injected into the host can be from multiple mechanisms, including: (1) photo-ionisation of gas up to CGM scales (e.g., \cite{arrigoni-battaia19,costa22}); (2) in-situ radiation pressure on gas and dust (located on CNR/torus scales to ISM scales; e.g., \cite{ishibashi18,costa18}); (3) powerful wide-angle winds associated with the accretion discs, which are likely radiatively-driven through electron scattering or through ultra-violet line driving (e.g., \cite{murray95,proga00,costa20,mizumoto21}); (4) collimated jets, which could efficiently couple to the gas on various spatial scales depending on power, inclination, ISM properties, etc. (e.g., \cite{mukherjee18,tanner22}; see Section \ref{sec:coupling}) and; (5) Compton (X-ray) heating in the region of the accretion disc, or even further out \cite{sazonov05,gan14} \footnote{We note that the inverse Compton effect could be important for cooling gas that is above the Compton temperature, including the effect of the AGN radiation field on the outflowing shocked gas, which could be crucial for establishing the potential impact of AGN-driven winds \cite{king15,bourne15}.}. Mechanisms (2)--(5) could all result in the propagation of mechanical energy into the host galaxies by shocking gas and/or by sweeping up material into an outflow (see Section~\ref{sec:gasimpact}). 

In the literature, the terms ``wind'', ``outflow'', and ``jet'' can be used interchangeably in a way that can cause confusion for interpretation. For AGN we encourage using the term ``wind'' to describe the launching mechanism associated with accretion discs and ``outflow'' to describe ISM/CGM material that has been swept up/entrained by the various driving mechanisms. The term ``jet'' may refer to the initial physical process (relativistic, collimated, beam of charged particles), or any other collimated outflowing structure, but this should be carefully defined in each study.

\subsection{Radio-identified AGN}\label{sec:radio}
We consider it important to provide a discussion on radio emission from AGN in the context of AGN feedback studies. The related terminology can cause some of the greatest confusion across both observational and theoretical works. 

The earliest radio observations were only sensitive to the most powerful extra-galactic sources which are clearly dominated by jets. The term `radio galaxy' is sometimes still used to define any galaxies considered to host radio-detected jets. However, in the modern era of radio astronomy, the origin of the detected radio emission can trace a wider range of processes (see below). The terms `radio loud' and `radio quiet' are also commonly used to classify AGN. Traditionally, radiatively efficient AGN are separated into these classes based upon the ratio of radio luminosity to an accretion luminosity (e.g., \cite{kellermann89,xu99,klindt19}). AGN that are dominated by radio emission above some threshold are classified as `radio loud'. However, `radio loud' is sometimes used more loosely in the literature to refer to galaxies where any radio emission associated with an AGN is identified, or to simply define if a galaxy is radio-detected. It is crucial to carefully define what is meant by a `radio galaxy' or `radio loud' AGN in each individual study, and to take caution in determining if the radio emission can, or can not, be unambiguously associated to radio jets. 

\begin{figure}[H]
\centering
\includegraphics[width=9.5cm]{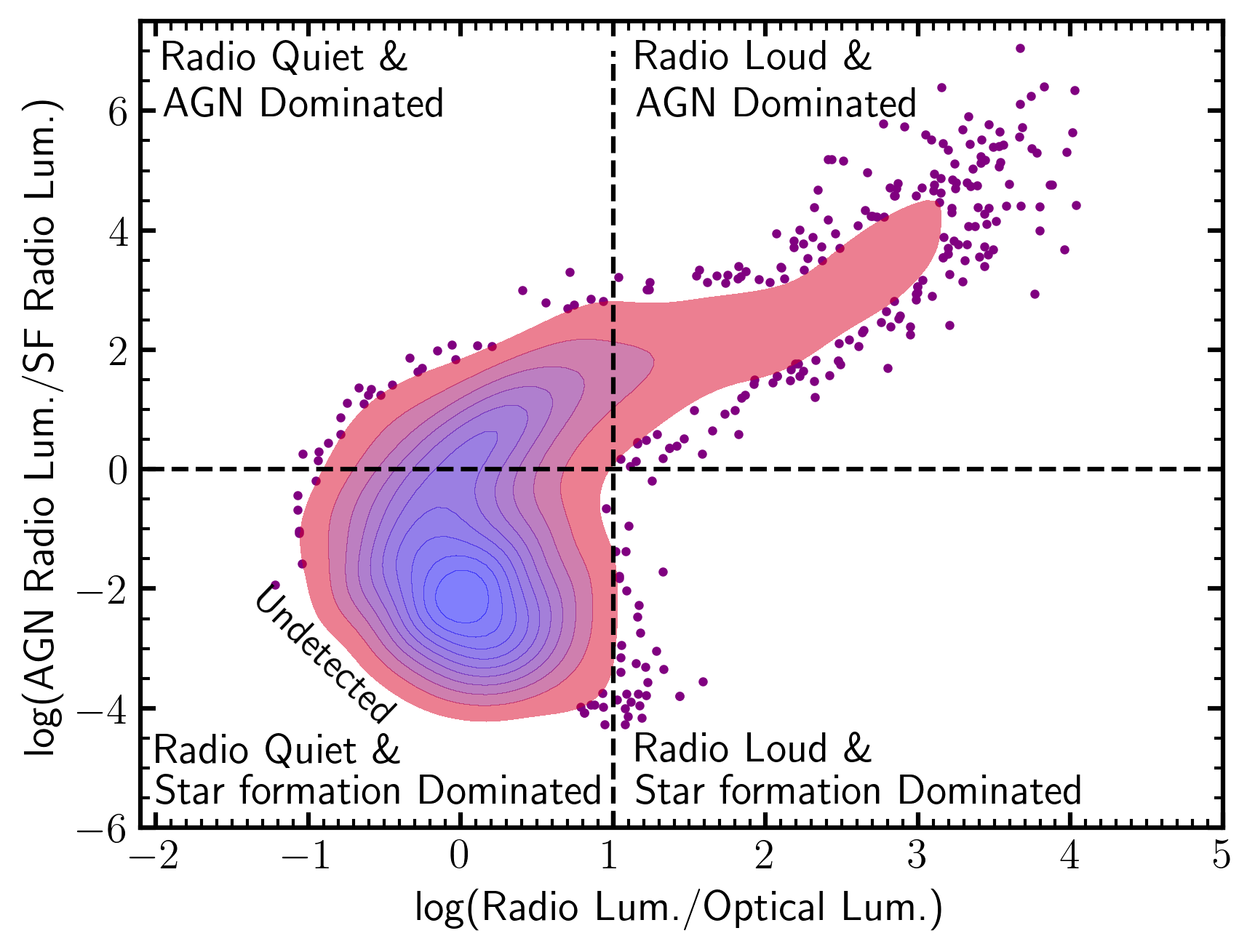}
\caption{Schematic representation of the data from Figure 12 in \cite{macfarlane21}, showing a comparison between the `radio loud' criteria for quasars (x-axis) (i.e., $R$ =$L_{\text{5GHz}}$/$L_{\text{4400\AA}}$, with a cut-off value at $R=10$) and the ratio of radio luminosity attributed to AGN to that attributed to star formation (y-axis). Sources below the detection threshold of the surveys would fill out the bottom-left of this diagram.
\label{fig:radio}}
\end{figure}

Due to the relationship between star formation rate (SFR) and associated radio luminosity in star forming galaxies, it is possible to estimate the expected contribution of star formation to the total observed radio luminosity (e.g., \cite{helou85,ivison10}). Therefore, any galaxy with significant radio emission above this is expected to have radio emission associated with an AGN, and are sometimes labelled as a `radio excess' galaxy/AGN or as a `radio AGN' (e.g., \cite{best12,delmoro13,macfarlane21}). In Figure ~\ref{fig:radio}, we provide a schematic representation of the difference between the traditional `radio loud' criteria and a radio excess criteria for radiatively efficient AGN. This figure is motivated by the data presented in Figure 12 in MacFarlane et~al. \cite{macfarlane21}, who study SDSS quasars with LOFAR observations.\footnote{Although the values in \cite{macfarlane21} are generated from a population model, with values not calculated for individual sources, they do represent the general observed quasar population trends. We also note, unlike in \cite{macfarlane21}, we do not assume that radio emission associated with AGN (as opposed to star formation), is necessarily attributed to jets.}

The traditional definition of `radio loud' effectively identifies galaxies with powerful jets and it is commonly accepted that $\sim$10--20\% of quasars are `radio loud' (Figure~\ref{fig:radio}, top-right quadrant; e.g., \cite{kellermann89,zakamska04,kellermann16,macfarlane21}). Nonetheless, {\em it does not necessarily mean that the radiative energy output is not energetically important for feedback}. A `radio loud' quasar is expected to host powerful accretion disc winds and have significant radiation pressure (e.g., \cite{king15,ishibashi18}; Figure~\ref{fig:powersources}). Furthermore, the `radio loud' definition misses the {\em majority} of the radiatively efficient AGN where the radio emission is AGN dominated (Figure~\ref{fig:radio}, top-left quadrant). 

A crucial point is that populations of radio-detected AGN contain a mixture of radiatively efficient and radiatively inefficient sources (see Figure~\ref{fig:powersources}). Radio-detected AGN, which exhibit weak or no optical emission lines are known as ``low excitation radio galaxies'' (LERGs) and appear to have little-to-no evidence of radiative emission associated with an accretion flow; this makes LERGs one candidate class of radiatively inefficient AGN (see review in \cite{hardcastle20}).\footnote{We note that ``Low-ionization nuclear emission-line regions'' (LINERs) are another class of galaxies, where the observed emission is not dominated by photoionisation from AGN, nor from star-forming regions and a subset of these sources may be associated with radiatively inefficient AGN (e.g., \cite{heckman80,flohic06,ho09}).} Radio-detected AGN that have also signatures of radiatively efficient accretion are known as ``High excitation radio galaxies'' (HERGs).

Whilst it is sometimes assumed that excess radio emission (above star formation) should be attributed to radio jets, this is not necessarily the case. For example, if AGN-driven outflows cause shocks in the ISM, this can be another significant contribution to the observed radio luminosity (e.g., \cite{nims15,zakamska16,panessa19}). Whilst diagnostics such as the `radio excess' criteria and high brightness temperatures (e.g., \cite{morabito22}) can be used to distinguish between AGN and star formation related emission, it remains observationally challenging to disentangle emission associated directly with a radio jet and that from outflow-induced shocks (e.g., \cite{jarvis21,fischer23}). Theoretical predictions will help guide appropriate diagnostics (e.g., \cite{bicknell18,meenakshi22}). 

In summary, radio emission associated with AGN events (both radiatively efficient and radiatively inefficient) is extremely prevalent and could even be ubiquitous (at least at low levels) for galaxies with the highest stellar masses (i.e., $\sim$10$^{11}$ solar masses; \cite{sabater19}). The origin of this radio emission is often ambiguous and detailed work is required to establish the underlying physical processes and the connection to AGN feedback.


\subsection{Connecting observed AGN populations to traditional `feedback modes'}\label{sec:energy_terminology}

We do not attempt to review AGN feedback models in this article. However, we make some brief notes about interpreting observed AGN populations in the context of some feedback models. Early semi-analytical models and hydrodynamic simulations invoked simple prescriptions for AGN feedback (e.g., \cite{springel05,sijacki07,hopkins06,bower06,croton06,somerville08,dubois10}), which have been influential on the community thinking about different `modes' of black hole growth and corresponding `modes' of AGN feedback (see discussion in e.g., \cite{bower12}). For example, the term `quasar mode' (`radiative mode', or `starburst mode') was considered important for rapid periods of growth (sometimes associated with mergers), with the idea that ISM could be expelled from the host galaxies. The term `radio mode' (or `hot-halo mode' , `maintenance mode', or `kinetic mode'), was considered important for lower accretion rate systems, and acts more through a regulation of cooling of the gas on large scales. Whilst appropriately simple at the time, the direct connection to {\em observed} populations of radio-detected AGN and quasars is not fully applicable with modern understanding. Furthermore, two distinct modes of feedback is unlikely to be the complete story.

We have established that during radiatively efficient AGN events, multiple mechanisms of energy input can result in significant {\em mechanical} forms of energy into the gas, through outflows driven by winds, jets etc. (Figure~\ref{fig:powersources}; Section~\ref{sec:energy}). Furthermore, as discussed in Section~\ref{sec:radio}, radio emission associated with AGN is prevalent for both radiatively efficient (high accretion rate) and radiatively inefficient accretion (low accretion rate) regimes. Therefore, we should avoid assuming that radio-detected AGN are universally associated with the traditional `radio mode' of feedback. It should also be avoided to assume that quasars provide only an `ejective' type of feedback on the ISM, and are only associated with accretion disc winds or radiation (cf. the traditional implements of a `quasar mode'). Quasars, and all radiatively efficient AGN, can also regulate gas cooling ('regulative' feedback), could have an important impact on ejection of gas out to CGM scales, and can have jets as an important energy injection mechanism (e.g., \cite{jarvis19,costa20,oppenheimer20,cicone21,audibert23}). 


\section{Energy-gas coupling: outflow kinematics and energetics}\label{sec:gasimpact}

In Section \ref{sec:demographics} we discussed the potential mechanisms by which AGN can inject energy into their host galaxies. However, it is important to remember that the relevance for AGN feedback requires this energy to couple to the multi-phase gas. Here we summarise observational results on multi-phase gas outflows based on samples of AGN (Section \ref{sec:outflows}), discuss different AGN and galaxy properties that might have a key impact on the energy-gas coupling (Section \ref{sec:coupling}), and describe how to constrain the outflow energetics and potential for impact from the observational point of view (Section \ref{sec:energetics}). 

\subsection{Multi-phase gas outflows}\label{sec:outflows}

Gas outflows, one of the clearest and hence most studied forms of AGN feedback, impact the ISM by removing, heating and/or mixing the gas (see Figure \ref{fig:TDscales}). Hundreds of observational studies report evidence of outflows observed in different gas phases, including the 1) hot ionised (T$_{\rm gas}\sim$10$^{5-8}$ K and n$_{\rm gas}\sim$10$^{6-8}$ cm$^{-3}$) using X-ray absorption lines \cite{longinotti13,tombesi13,chartas21}; 2) warm ionised (T$_{\rm gas}\sim$10$^{3-5}$ K and n$_{\rm gas}\sim$10$^{2-5}$ cm$^{-3}$), mainly through emission lines detected in the rest-frame UV \cite{murray95b,sulentic07,coil11,martin15}, optical \cite{harrison14,villar16,forster14,forster19,rose18,scholtz18,scholtz20} and near-infrared \cite{rupke13,ramosalmeida19,speranza22,riffel23}; 3) neutral atomic (T$_{\rm gas}\sim$10$^{2-3}$ K and n$_{\rm gas}\sim$10$^{1-2}$ cm$^{-3}$) by means of the H~I absorption line \citep{morganti05,morganti16}, the sodium doublet absorption (NaID; \cite{rupke05,cazzoli16,concas19} and the [C II] fine-structure emission \cite{maiolino12,bischetti19}; and 4) molecular (T$_{\rm gas}\sim$10$^{1-3}$ K and n$_{\rm gas}>$10$^3$ cm$^{-3}$). The latter phase can be probed, for example, using the rotational and vibrational H$_2$ lines in the near-infrared (hot molecular; \cite{rupke13,ramosalmeida19,riffel23}), the rotational lines in the mid-infrared (warm molecular; \cite{dasyra11,riffel20}), the hydroxyl (OH) transitions in the far-infrared \cite{sturm11,veilleux13,spoon13,gonzalez17} and the carbon monoxide (CO) emission lines in the (sub)millimeter \cite{feruglio10,cicone14,pereira18,fluetsch19,lamperti22,ramosalmeida22,scholtz23}. These spectral features trace outflows on different spatial scales, that go from sub-parsec in the case of the hot ionised phase (i.e., BLR scales), and from hundreds of parsecs to $\sim$10 kpc for the ionised and molecular gas phases (see Figure \ref{fig:TDscales}).

Despite the wealth of observational effort invested in characterising AGN-driven outflows, the overall majority of studies report single-phase estimates of their properties that provide an incomplete view of the AGN feedback phenomenon \cite{cicone18}. Accurate outflow properties in different gas phases need to be determined in representative AGN samples to estimate how much each phase contributes to the overall mass and energy budget. From the few studies reporting multi-phase outflow properties for samples of AGN (e.g., \cite{rupke17,fluetsch21,riffel23,speranza24}), the cold molecular outflows would carry the bulk of the mass, followed by the neutral and the warm ionised. In the local Universe, the molecular outflows are slower (v$\lesssim$300-500 km~s$^{-1}$) and more compact (r$\lesssim$1-2 kpc) than the ionised outflows ($\sim$500--1000 km~s$^{-1}$; r$\sim$1-10 kpc). In the high redshift Universe (at cosmic noon, i.e., z$\sim$1--3, and earlier), most of the works reporting outflow measurements focus on the ionised gas phase, since the bright optical emission lines are shifted to the infrared. Integral field spectroscopy of quasars with different radio luminosities revealed evidence of fast ionised outflows on galaxy scales \cite{nesvadba08,carniani15,kakkad20,scholtz20,vayner21,concas22,wylezalek22,vayner24}. The neutral atomic and molecular gas phases of the outflows have also been studied at high redshift, and although some works report the presence of massive, galaxy-scale high-velocity outflows in luminous quasars \cite{maiolino12,cicone15,bischetti19,stacey22}, overall there is no evidence for them being a widespread phenomenon, at least those probed by the [C~II] emission line \cite{decarli18,novak20}. A handful of studies on individual galaxies reported multi-phase outflow measurements, mainly targeting the cold molecular and ionised phases \cite{herrera19} but also the ionised and neutral atomic \cite{deugenio23}. \cite{vayner21b} characterised the ionised and cold molecular outflows in four quasars at z$\sim$2 with powerful jets. They found that, although most of the outflow mass is in the molecular phase (which is clumpier and more compact), the kinetic energies are higher in the ionised phase. 

With a few exceptions, the multi-phase outflow masses derived from observations are modest, but observational limitations might be preventing us from detecting the most extended and diffuse component of the multi-phase outflows, which might carry the bulk of their mass and/or energy. For example, in the case of the cold molecular gas, this diffuse component might be missed by interferometric observations that do not include short baselines \cite{cicone18}. Deep X-ray observations have revealed the presence of hot X-ray-emitting gas (T$_{\rm gas}\sim$10$^{6-7}$ K) extending to CGM scales in quasars (e.g., \cite{veilleux14,greene14,lansbury18}), and diffuse, extended gas on scales of up to $\sim$50 kpc have been reported for neutral hydrogen around radio galaxies (e.g., \cite{morganti13}). Some of this diffuse emissions might be participating in, or being entrained by, the outflows. Therefore it would largely increase the outflow mass, and in the case of the hot ionised gas, its energy. Whilst we wait for new a generation of X-ray observatory, that will be required to fully characterise the hottest gaseous phases \cite{russell23}, indirect approaches will need to be taken \cite{simionescu19,brownson19,richter20,richings21}.

From the observational point of view 
it is not clear yet whether the different outflow phases are different faces of the same phenomenon, with certain phases transitioning to others \cite{costa18}, or just unrelated events. Deep, spatially resolved multi-phase outflow measurements of AGN of different luminosities, hosted in galaxies with different properties and environments, are necessary for advancing our understanding.    

\subsection{Coupling between energy and gas}\label{sec:coupling}

\begin{figure}[H]
\centering
{\par
\includegraphics[width=6.55cm]{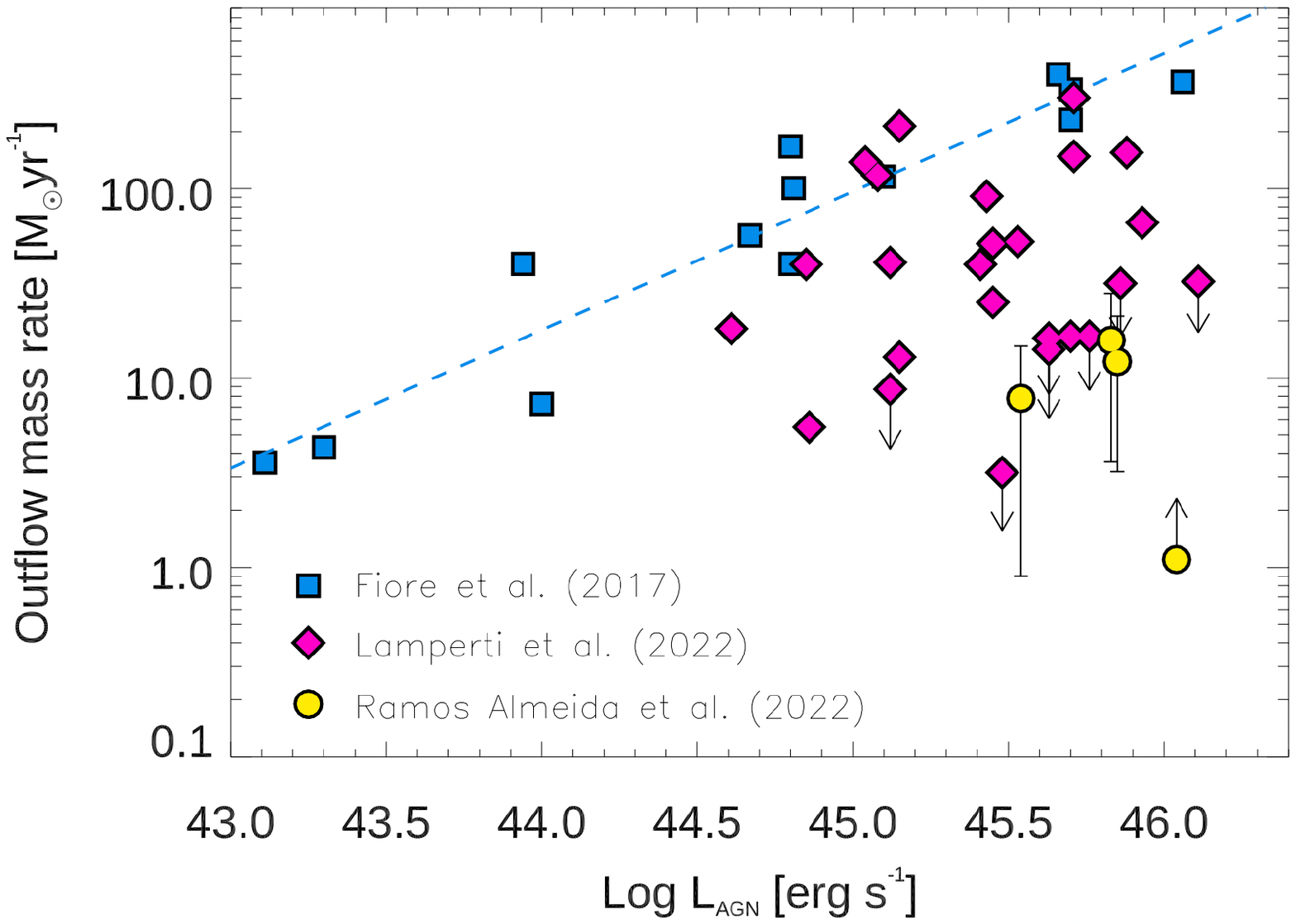}
\includegraphics[width=7.2cm]{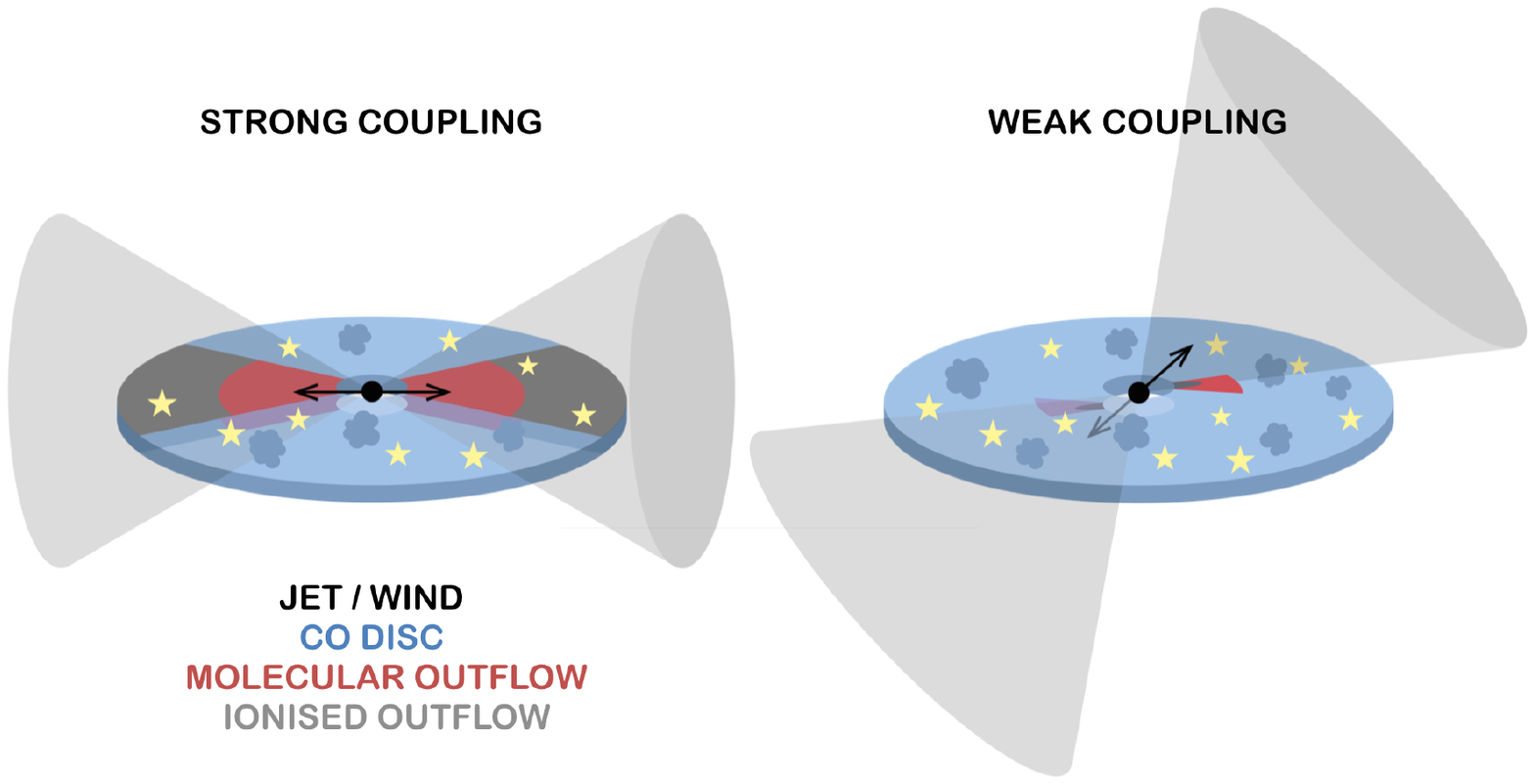}\par}
\caption{Molecular mass outflow rate versus AGN luminosity (left panel). The dashed line is the fit to the data compiled by Fiore et al.~\cite{fiore17}. Diamonds are the ULIRGs from Lamperti et al.~\cite{lamperti22} and circles the type-2 quasars from Ramos Almeida et al.~\cite{ramosalmeida22}. The plots on the right are examples of a strong (weak) coupling scenario, where the jet/wind/ionised outflow subtends a small (large) angle with the molecular gas disc, launching a massive (modest) molecular outflow, shown in red. Credit (right panel): Ramos Almeida, A\&A, 658, A155 (2022) \cite{ramosalmeida22}, reproduced with permission ©ESO.}
\label{fig:coupling}
\end{figure}

According to observational scaling relations found from outflow measurements of AGN in the local Universe, the more luminous the AGN the more massive and faster the gas outflows that it drives (e.g., \cite{cicone14,fiore17}). However, these scaling relations were derived from observations of CO-bright targets known to have strong multi-phase outflows (e.g., Mrk\,231 \cite{feruglio15}, NGC\,1068 \cite{garciaburillo14,saito22}, IC\,5063 \cite{tadhunter14,morganti15}), some of them hosted in ultra-luminous galaxies (ULIRGs). More recent works have now started to populate the AGN luminosity-outflow mass rate plane with less biased AGN and ULIRG samples \cite{ramosalmeida22,lamperti22,speranza24}, and the linear relation vanishes (see left panel of Figure \ref{fig:coupling}). This indicates that having a high AGN luminosity does not guarantee the presence of a massive outflow, as the coupling between energy and gas depends on other factors that we mention below.

\vspace{-0.5cm}

Aside from AGN luminosity, a key property that determines the level of coupling between energy and gas is orientation. When a jet and/or disc wind subtend a small angle relative to the gas disc, the latter is entrained and the jet/wind energy is transmitted more efficiently (strong coupling scenario in Figure \ref{fig:coupling}) than when the angle is large (weak coupling scenario). Something similar can happen when an ionised outflow is coplanar or quasi-coplanar with a galaxy disc: the molecular gas is then entrained by the ionised outflow, becoming a molecular outflow, normally directed along the kinematic minor axis \cite{garciaburillo21,ramosalmeida22}. 
The importance of orientation has been shown by hydrodynamic simulations of jet-ISM interactions: jets almost coplanar with the galaxy discs (strong coupling scenario in Figure \ref{fig:coupling}) entrain the gas more efficiently than jets subtending a larger angle (weak coupling scenario), inducing higher outflow velocities and disturbance \cite{mukherjee18,meenakshi22,talbot21,talbot22,audibert23}. Jets have been proposed to be responsible for the enhancement in velocity dispersion in the direction perpendicular to the jet observed in AGN of different bolometric luminosities with modest jet powers \cite{couto17,balmaverde19,balmaverde22,venturi21,girdhar22,ramosalmeida22,venturi23}. Another factor with potential influence on the coupling between jets and ISM is jet power. Simulations show that low-power jets (P$_{\rm jet}\lesssim$10$^{45}$ erg~s$^{-1}$) get trapped for a longer time by the ISM and disrupt the surrounding gas over a larger volume, and for a longer period, than jets of higher powers (P$_{\rm jet}\gtrsim$10$^{45}$ erg~s$^{-1}$), which are capable of ``drilling'' through the ISM \cite{nyland18}. Therefore, the low-power jets often detected in radiatively-efficient AGN such as Seyferts and quasars can be more disruptive and induce more massive outflows than high power jets \cite{harrison15,morganti15,morganti17,girdhar23,cresci23,veilleux23}. Finally, the distribution of molecular gas is another factor leading to more or less efficient coupling. Some of the most powerful molecular outflows are found in ULIRGs \cite{cicone14,feruglio10}, associated with galaxy mergers and high nuclear column densities. The latter might favor the launch of massive molecular outflows by increasing the coupling between the jets/winds and the ISM \cite{wagner16,costa20,venanzi20}. It has been proposed that this dust-enshrouded phase is followed by an active feedback phase (blow-out phase) \cite{alexander12,glikman12,banerji15} that modifies the nuclear molecular gas distribution, creating cavities that have been predicted by hydrodynamic simulations of massive galaxies with disc winds \cite{mercedesfeliz23} and observed in nearby AGN using ALMA \cite{rosario19,feruglio20,garciaburillo21} (see Section \ref{sec:localimpact}).

Observations of representative samples of AGN of different bolometric and radio luminosities, hosted in galaxies of different masses and morphological types, are needed to understand the interplay between the different outflow gas phases and the impact of the different factors listed above on the coupling between the outflows and the ambient gas. This requires observing proposals in different telescopes that are expensive in terms of observing time and that can be considered ``incremental'' in terms of science goals, but we argue that they are necessary to move forward in AGN feedback studies.  


\subsection{Outflow energetics and potential for impact}\label{sec:energetics}

Here we consider gas outflows as any gas that is not just photoionised, excited and/or shocked by the AGN, but also kinematically disturbed by it, as in \cite{davies20}. Identifying outflows through emission line profiles is challenging because different gas components across the host galaxy having different kinematics contribute to the shape of the line profiles: rotation, stripped material, infalling gas, companion objects, etc (see e.g., \cite{concas22}). The most widely used method consists of assuming that each kinematic component can be described by a Gaussian (i.e., parametric methods). In the simplest case scenario, the emission lines can be described by the sum of a narrow component ($\lesssim$400 km~s$^{-1}$) tracing virial motions, and another component(s), generally but not necessarily broader and shifted relative to the systemic velocity, associated with outflowing gas. Using parametric methods it is possible to measure the properties of each kinematic component, but in galaxies with complex kinematics it is difficult to ascribe a physical meaning to each of them \citep{bessiere22,hervella23}. In the latter case, non-parametric methods might be more appropriate \cite{whittle85,harrison14,speranza2021} because they make it possible to identify high-velocity gas by measuring emission line velocities at fixed fractions of either the peak or integrated line flux (see Figure 4 in \cite{harrison14}). We refer the reader to Hervealla-Seoane et al.~\cite{hervella23} for an analysis of the influence that using different parametric and non-parametric methods have on the outflow kinematics and energetics. 

In the case of the molecular gas, often studied using the CO emission lines detected in the (sub)-mm regime, outflows usually have lower velocities than their ionised counterparts \cite{vayner21b,ramosalmeida22,speranza24}, making it even more difficult to disentangle them from the virial motions around rotation. Therefore, unless clear emission line wings are detected (e.g. \cite{cicone14,audibert19}), it is not possible to identify any detected non-circular motion with radial outflows without a careful analysis of the gas kinematics. The observed moment maps and position velocity diagrams can be compared to simple rotating disc models in order identify molecular outflows \cite{garciaburillo14,dominguez21,ramosalmeida22}. Once identified, the outflow flux can be calculated using more or less conservatives approaches \cite{audibert23}. 
Regardless of the method used to characterise the gas kinematics, several observational effects including the spectral response and resolution of the instrument, beam smearing, projection, etc. need to be considered and accounted for in order to reliably characterise the outflow properties. These effects, their implications and ways of mitigation have been largely discussed in the literature \cite{carniani15,Husemann16,harrison18}.


To infer the potential impact of the outflows on the host galaxies we need to measure outflow physical properties such as mass, mass rate, and kinetic power (M, \.M, \.E$_{\rm kin}$). This requires estimates of other quantities such as the outflow density (see \cite{harrison18} for discussion on different methods and caveats) and metallicity, and to assume an outflow geometry. The latter can be spherical, (multi-)conical or a slab/shell of certain radial thickness, and it also has an impact on the derived outflow mass rates \cite{lutz20}. Other assumptions that are often made are that the outflow clouds have the same density and metallicity and no acceleration or deceleration of the outflowing gas. In the case of the cold molecular outflows estimated from CO emission lines, an $\alpha_{\rm CO}$ factor needs to be assumed to convert from measured carbon monoxide luminosities to estimated molecular hydrogen (H$_2$) masses. Values of $\alpha_{\rm CO}\leq1$ are usually measured/assumed for molecular outflows \cite{morganti15,garciaburillo14,ramosalmeida22}, which is significantly lower than the Galactic factor ($\alpha_{\rm CO}$=4.35; \cite{bolatto13}).

The outflow mass rates can be compared with the SFR to derive the mass loading factor ($\eta$=\.{M}/SFR) and hence have an estimate of the gas mass that is being evacuated by the outflow compared to that being consumed by star formation. The trouble is that often the spatial scales corresponding to the outflow (few kiloparsecs at most) and to integrated SFRs (galaxy scales) are different, and therefore the mass loading factors might be misleading \cite{ramosalmeida22}. The outflow momentum rate (\.M$\times$v) divided by the AGN radiation momentum rate (L$_{\rm bol}$/c) can provide information about whether the outflows are energy or momentum driven \cite{king15,fiore17}, and the kinetic power (\.E$_{\rm kin}$) is usually compared with either the jet power or the AGN bolometric luminosity to have an idea of whether the kinetic power of jets and/or AGN radiation can drive the outflow \cite{harrison14,audibert23,venturi23,speranza24}. However, many factors compound to make it observationally challenging to obtain reliable measurements of bolometric luminosity. The required bolometric correction factors, to convert from a luminosity in one wavelength regime to the total AGN luminosity, remain challenging and uncertain \cite{Netzer19}. Furthermore, a dominant fraction of AGN may be heavily obscured and can even be completely missed in many surveys \cite{hickox18}. Finally, as previously discussed in Section~\ref{sec:overview}, accretion rates and the resulting energy output can be highly variable, on temporal scales much shorter than the outflow dynamical timescales. A significant challenge also exists when attempting to determine the available energy from jets. Scaling relationships have been created between observed radio luminosities and jet powers, using measurements of the mechanical work inferred from the jet-induced cavities inside X-ray emitting atmospheres of massive galaxy environments (e.g., \cite{merloni07,birzan08,cavagnolo10}). These relationships are used extensively in the literature to convert a radio luminosity into the available power from jets as a potential feedback mechanism. However, there are many systematic uncertainties in these relationships, even for the regimes (i.e., jet powers, galaxy environments) from which they were calibrated (e.g., \cite{kokotanekov17,croston18}). Crucially, it is not-at-all clear if they can be extended to different galactic environments, or to different regimes of jets. For example, lower power jets, which are orientated into a gas rich ISM can quickly be frustrated/disrupted and the observed radio luminosity may be a poor tracer of jet power (e.g., \cite{mukherjee18,tanner22}). Furthermore, the observed radio luminosity may be the result of many processes, and not simply a direct tracer of jets (\cite{panessa19,fischer23}; Section~\ref{sec:radio}). Therefore, when using observed AGN luminosities and jet powers to test theoretical predictions, or to establish the dominant energy driving mechanisms, it is crucial to consider all of these systematic uncertainties. 

The ratio between the outflow kinetic power and AGN luminosity is often referred to as the kinetic coupling efficiency and it has been widely used by the community in an attempt to compare observational results with the predictions from cosmological simulations. In the seminal work of Di Matteo et al.~\cite{dimatteo05}, the feedback efficiency ($\epsilon_f$; the fraction of AGN luminosity that couples to the gas in the vicinity of the black hole) was calibrated to 5\%, so the local M$_{\rm BH}$-$\sigma$ relation was reproduced by the simulation. As widely discussed in \cite{harrison18}, the comparison between the observed kinetic coupling efficiencies (\.E$_{\rm kin}$/L$_{\rm bol}$) and the $\epsilon_f$ values adopted in cosmological simulations is not straightforward. First, in the case of radiation-driven gas outflows, the feedback efficiencies are calibrated (to $\sim$0.005-0.15 in more recent simulations; see \cite{harrison18} and references therein) and hence are not a prediction themselves \cite{bourne15,crain15,schaye15}. Second, only a fraction of the injected energy will be transformed into outflow kinetic energy depending on the ISM properties, the gravitational potential, etc. Therefore, only a small kinetic coupling efficiency may be actually measured from observations. Finally, as several observational studies show, outflows can be launched/accelerated by compact jets, even in radiatively efficient AGN, challenging how useful a comparison to the bolometric AGN output would be in these cases (see Sections \ref{sec:energy_terminology} and \ref{sec:outflows}). 
As discussed in Sections \ref{sec:localimpact} and \ref{sec:longterm}, there are better ways to compare observations with the predictions from simulations than just taking a fiducial theoretical value as a reference to determine whether kiloparsec scale outflows are relevant from an energetic point of view. 


\section{Localised and transient impact on gas and star formation}\label{sec:localimpact}

According to cosmological simulations, it is the cumulative output of several AGN events that are most relevant for the global and long-term impact on their host galaxies (see Section \ref{sec:longterm}). The current AGN state of a galaxy is therefore not a useful proxy for studying the global and long-term impact of AGN feedback on gas and star formation, at least in the local Universe (see \cite{silk24} for a recent discussion of the potential impact of AGN feedback across cosmic time). However, by studying currently active AGN, using spatially-resolved observations, we can obtain crucial information on the localised impact of AGN feedback (labelled as `3' in Figure~\ref{fig:overview}), which is needed to understand how the energy couples with the gas, and under which circumstances it enhances or reduces star formation efficiency.



As already mentioned in Section \ref{sec:overview}, AGN should be treated as events or episodes that are much shorter than the typical timescales of star formation. In the case of the outflows, the dynamical timescales are usually $\sim$1-10 Myr depending on their velocity and extent (for example, for an unimpeded outflow expanding at an average velocity of 1000 km~s$^{-1}$ it takes one million years to reach a radius of one kiloparsec). Therefore, if we aim at investigating the impact, either positive or negative, that the present outflows have on star formation, we should be looking at young stellar populations (YSPs), dominated by O and B stars. The latter can be traced using stellar population modelling of rest-frame UV/optical spectra \cite{bruzual03} and also the polycyclic aromatic hydrocarbon (PAH) features detected in the rest-frame mid-infrared \cite{peeters04}. In the case of other star formation tracers such as H$\alpha$, [Ne II], and the 24 $\mu$m flux it is difficult to disentangle the contribution from star formation and AGN (e.g., \cite{stanley18,scholtz21,lamperti21}), although it has been argued that PAHs could also be excited by AGN radiation \citep{howell07,jensen17}. Stellar synthesis modelling is challenging in AGN because in the case of type-1 AGN the strong continuum dilutes the stellar absorption features. Type-2 AGN are better suited for this type of analysis, although scattered AGN light from the obscured nucleus can significantly contribute to the UV/optical continuum, and therefore can be interpreted as a YSP  \cite{bessiere17}. This degeneracy can be broken if high order Balmer lines (e.g., H9, H10, H11 and H12; 3835--3750 \AA) are detected, since they are tracers of young stars. It is also important to account for the nebular continuum coming from the free–free, free–bound and two photon decay continua associated with the gas nebula \cite{dickson95} to accurately characterise the YSPs. 

Evidence for recent star formation in AGN based on the detection of PAH features has been reported from kiloparsec scales \citep{zakamska08,zakamska16,deo09,diamond12} to as close as tens of parsecs from the AGN \citep{sales13,alonsoherrero14,esquej14,ramosalmeida14,esparza18}. However, it is important to remember that AGN radiation and shocks might modify the structure of the aromatic molecules and/or destroy the smallest grains, hence suppressing the short-wavelength features (6.2, 7.7, and 8.6 $\mu$m; \cite{smith07,diamond10}). Luckily the larger and neutral molecules that produce the 11.3 and 17 $\mu$m features are more resilient and they have been found to be enhanced in AGN relative to the short-wavelength PAHs, at least in Seyfert galaxies \citep{garciabernete22}. At higher AGN luminosities it is not clear if even the largest PAH molecules can survive the AGN radiation field \cite{martinez19,xie22,ramosalmeida23}, but upcoming JWST datasets will be crucial to study the viability of PAH features as tracers of star formation in AGN of different luminosities, at different distances from the AGN, and in and out the outflow/jet regions \cite{garciabernete22b,lai22,lai23}.

AGN-driven outflows and jets can trigger star formation locally by compressing the surrounding gas, as it has been shown theoretically (e.g., \cite{silk13,zubovas17,mercedesfeliz23,raouf23}) and observationally using integral field spectroscopy and/or interferometry (e.g., \cite{cresci15,salome17,carniani16,shin19,bessiere22}). Mercedes-Feliz et al.~\cite{mercedesfeliz23} used the FIRE simulations incorporating hyper-refined disc winds to study the impact of quasar outflows with kinetic powers of 10$^{\rm 46}$ erg~s$^{-1}$ on the circumnuclear disc (CND; scales of $\sim$100 pc; see Figure \ref{fig:TDscales}) of massive galaxies at z$\sim$2. These outflows drive the formation of a central gas cavity and increase the local star formation efficiency at its inner edges, where gas is being compressed. Despite this localised positive feedback, the simulations show that the outflows reduce the surface density of star formation across the galaxy (i.e., negative feedback) by limiting the availability of gas rather than evacuating it \cite{torrey20,mercedesfeliz23}. Recent observational work using data from legacy integral field spectroscopic surveys report evidence for suppressed gas fractions and SFRs in the central kiloparsec of galaxies hosting AGN as compared to matched controls (e.g., \cite{sanchez18,ellison21,lammers23}).

The creation of molecular gas cavities of tens to hundreds of parsecs by AGN-driven outflows and/or jets is one of the most clear observational evidences of the direct impact of AGN feedback on the host galaxies \cite{rosario19,feruglio20,garciaburillo19,garciaburillo21}. Using ALMA CO observations at $\sim$10 pc resolution of a sample of nearby AGN with X-ray luminosities in the range $\sim$10$^{39-43}$ erg~s$^{-1}$, García-Burillo et al.~\cite{garciaburillo21} found that the most luminous AGN within the sample show evidence for molecular outflows and smaller molecular gas concentrations in the inner r$\sim$50 pc (torus scales) of the galaxies relative to the inner r$\sim$200 pc (CND scales) than the less luminous AGN in the sample. They interpreted these results as the imprint of feedback on the nuclear distribution of molecular gas: at low AGN luminosities, galaxies build their nuclear molecular gas reservoirs via accretion of gas, and as accretion and subsequent feedback become more efficient, jets and winds push molecular gas outwards, producing the cavities predicted by hydrodynamic simulations (e.g., \cite{torrey20,mercedesfeliz23,raouf23}). These molecular gas cavities, which have been detected in low-J CO transitions such as CO(1-0) \cite{ruffa22}, CO(2-1) \cite{rosario19,feruglio20,garciabernete21,ramosalmeida22}, and CO(3-2) \cite{garciaburillo19,garciaburillo21}, are filled with ionised and warm molecular (H$_2$) gas \cite{rosario19,garciaburillo19,feruglio20}, indicating that the AGN is modifying the properties of the cold molecular gas on CND scales. Indeed, for AGN observed in more than one CO transition, such as NGC\,1068 \cite{garciaburillo19}, NGC\,3100 \cite{ruffa22} and the Teacup \cite{audibert23} higher gas excitation is found in the central regions of the galaxies. All these galaxies have relatively low jet-powers (P$_{\rm jet}\sim$10$^{\rm 43}$ erg~s$^{-1}$) with different inclinations relative to the molecular gas discs, which might be responsible, together with AGN radiation and outflows, for the high gas excitation in specific regions of the galaxy centers.  

In summary, spatially-resolved studies of nearby AGN are revealing the complex interplay between AGN and the surrounding gas. Obtaining accurate measurements of this interplay, including outflow energetics and ambient gas and recent star formation properties, is key to revise the recipes used in simulations. The high angular resolution and sensitivity of facilities such as ALMA and the JWST now allow us to obtain spatially-resolved constraints at high redshift (e.g., \cite{deugenio23,tripodi24,ubler24}). 

\section{Global and long-term cumulative impact of AGN feedback}\label{sec:longterm}

To assess the relevance of AGN feedback in a cosmological context, requires statistical studies that search for the imprint of feedback on the overall galaxy population (labelled `4' in Figure~\ref{fig:overview}). 

Some of the best evidence that AGN have a global and long-term impact on galaxy evolution comes from the observation that AGN jets can offset cooling in dense galactic environments. Hence, they can regulate the rate at which gas can cool into galaxies and form stars. X-ray cavities excavated by radio jets are well studied in local cluster and group environments, and provide direct indication of the work done as these jet-induced lobes expand into the ICM/IGrM \cite{boehringer93,birzan08,mcnamara12,hlaveckLarrondo22}. This process, appears to provide the required amount of heating to balance cooling and may apply for massive early type galaxies in a wider range of environments and out to $z\sim$1 \cite{best06,danielson12,hlavaceklarrondo15,smolcic17,mcdonald18,hardcastle19}. In broad terms, there appears to be convincing evidence that the cumulative effect of AGN (see Figure~\ref{fig:overview}) is able to regulate cooling in massive systems, at least in the densest environments and the latest cosmic times. Nonetheless, some of these calculations remain uncertain, it is still an open question on how the energy is communicated to the ICM/IGrM, and the universal applicability across different environments and accretion states remains a matter of ongoing study \cite{birzan17,werner19,hardcastle20,bourne23}. Furthermore, whilst this could be an effective mechanism for maintaining low star formation levels, it may be insufficient to shut down, i.e., `quench', high star formation levels, which may require a more ejective form of feedback (e.g., \cite{dubois10,mccarthy11,davies20,zinger21}). 

One common type of study has been to measure the SFRs, gas fractions, and star formation efficiencies in samples of AGN host galaxies (see \cite{harrison17,ward22}, and references there-in). Some of these studies look for instantaneous impact of the observed AGN events on either the global (galaxy-wide) fuel for star formation (the molecular gas content) or directly on the SFRs. The two broad approaches are to compare these measured properties to matched samples of inactive galaxies (e.g., \cite{scholtz18,kirkpatrick19,circosta21,bischetti21,ramosalmeida22,molyneux24}), or to explore these properties as a function of AGN luminosity or the prevalence of AGN-driven outflows (e.g., \cite{rosario13,stanley15,balmaverde16,woo17,stanley17,wylezalek18,shangguan19,zhuang21,kim22}). There has been some evidence (although not a consensus) that luminous AGN at cosmic noon have moderately depleted levels of molecular gas (e.g., \cite{perna18,circosta21}). This potentially indicates that this gas has been excited to higher transitions (see Section~\ref{sec:localimpact}) or has been ejected. However, despite the wide range of approaches, and the different AGN samples, the general consensus is that radiatively efficient AGN typically reside in gas-rich, star-forming galaxies. This general picture is also consistent with statistical studies that look at the galaxy population as a whole; although not {\em all} radiatively efficient AGN live in star-forming galaxies, this is more likely to be the case for the most luminous AGN (e.g., \cite{aird19,grimmett20}). In contrast, radiatively inefficient AGN are more likely to be found in quenched (non star-forming) galaxies (e.g., \cite{miraghaei17}).

A comparison to cosmological simulations shows broad agreement with the above observational result, that high accretion rate AGN typically reside in gas rich, star-forming galaxies \cite{scholtz18,ward22,piotrowska22,bluck23a}. This likely tells us something about the shared fuel supply and feeding processes needed to invoke a luminous AGN event. Furthermore, it shows that these type of observations do not rule out AGN feedback models that are invoked in simulations. Indeed, a measured instantaneous radiative luminosity of an AGN is likely unconnected to the eventual global, cumulative, feedback on the host galaxy because an instantaneous AGN luminosity does not constrain the total (cumulative) energy injection that could occur during multiple periods of black hole growth (see Figure~\ref{fig:overview}). This all suggests that molecular global gas content and SFRs of statistical samples of galaxies hosting radiatively efficient AGN, are not expected to reveal strong signatures of galaxy-wide (global) AGN feedback (also see \cite{scholtz18,ward22,piotrowska22,bluck23a,bluck23b}). Instead, the global evidence for AGN feedback may be found in the wider galaxy population, which do not necessarily currently host an observable AGN event (e.g., \cite{martin18,martin21}). 

AGN feedback is largely required in galaxy evolution theory to explain the build up of quenched galaxies. Therefore, another approach to investigate the role of AGN feedback is to consider the galaxy population as a whole, and ask what is the most important factor/property for a galaxy to be quenched. The total black hole mass can be considered a tracer of the historic record of the total radiative energy output by AGN (see Figure~\ref{fig:overview}), because the majority of black hole growth occurs during radiatively efficient phases. Work has suggested a strong connection between black hole mass and quiescence in galaxies, for stellar-mass matched samples, with broad agreement between both observations and simulations (at least for massive central galaxies, as opposed to satellite galaxies; \cite{terrazas16,piotrowska22,bluck23a,bluck23b}). These works imply that it is the integrated (cumulative) history of energy injection from AGN that is the most important for suppressing star formation, and not the instantaneous AGN luminosity. Nonetheless, caution must be taken when comparing these observations to simulations that invoke effective feedback prescriptions that are only turned on above a certain black hole mass (also see \cite{ward22}). Furthermore, significant energy output may occur during AGN events that do not contribute significantly to black hole growth, for example from jets during radiatively ineffecient AGN events. 

Indeed, an important difference between some feedback models in cosmological simulations, is if the feedback is most efficient during periods of rapid black hole growth (cf. a `radiatively efficient' regime), or periods of low accretion (cf. a `radiatively inefficient' regime). In the former case the total energy output will be related to the black hole mass, whilst for the later this is not necessarily the case. Voit et~al.~\cite{voit24} show that the failure of one particular cosmological simulation to reproduce the observed black hole mass to halo mass relationship can be attributed to the fact that the implemented {\em efficient mode of feedback} for lifting baryons (which results in the star-formation quenching) occurs only during periods of {\em low accretion rates}. This appears to be in contrast to observations that suggest the majority of black hole growth occurs during the quenching process \cite{chen20,voit24}. Similar comparisons between observations of the overall galaxy population and simulations, will be important to further test and refine the different prescriptions of feedback models in simulations (e.g., \cite{habouzit21,kondapally23}).  Spatially-resolved measurements of star formation, and stellar populations, on large samples of galaxies in different environments can shed further insight into the quenching process and the potential role of AGN across different populations (e.g., \cite{bluck20,lammers23}).

Finally, we note that the effects of AGN feedback may be imprinted on CGM's surface brightness profiles of emission-line regions and on the distribution of metals (e.g., \cite{costa22,obreja24}). Different prescriptions of feedback result in different predicted properties of the CGM (e.g., \cite{davies20,costa22,khaire23,wright24}). This remains a promising route to further test feedback prescriptions and bridge the gap between theoretical predictions and observational studies. 

\section{Concluding remarks}

After decades of work, we draw the conclusion that there are no observations which are {\em inconsistent} with the broadest theoretical idea that AGN feedback is a crucial component of galaxy formation theory for explaining the properties of massive galaxies.  However, obtaining a complete physical picture of how this process works in the real Universe is an on-going challenge. It is a multi-scale problem over many orders of magnitude in spatial, temperature, and time scale (see Figure~\ref{fig:TDscales} and Section~\ref{sec:overview}). Therefore, it is inescapable that multiple observational approaches are required to tackle this problem, using a range of facilities and a breadth of galaxy samples, for addressing the many components of the overall process (see Figure~\ref{fig:overview}). Some key takeaway points from our holistic review are:
\begin{itemize}

    \item AGN should be considered {\em events} and not {\em objects} that persist in time. AGN are self-regulatory and variable; therefore, it may be difficult to directly relate a single accretion episode to a significant, global impact on galaxy properties. Ultimately, the properties of a galaxy will be influenced by the cumulative output of multiple accretion episodes/feedback events (see Section~\ref{sec:overview}). 
   
    \item Both high accretion rate (`radiatively efficient') and low accretion rate (`radiatively inefficient') AGN can have multiple, overlapping mechanisms for injecting energy into their hosts (see Figure~\ref{fig:powersources}). These can contribute to both ejective and regulative channels of feedback (see Figure~\ref{fig:overview}). Radio emission can trace a range of feedback mechanisms over a range accretion rates (see Figures~\ref{fig:powersources} and~\ref{fig:radio}). Therefore, care should be taken when comparing simplified theoretical feedback modes with observed AGN populations (see Section~\ref{sec:demographics}).  
   
    \item An AGN-driven outflow is gas that has been kinematically disturbed by a variety of possible driving mechanisms including accretion disc winds, radiation pressure, and jets. For the outflows to be relevant for feedback their energy has to couple to the multi-phase gas, and this coupling depends on several factors including AGN luminosity or jet-power, jet/wind orientation and ISM properties. More efficient coupling will result in a more significant impact on the properties/distribution of the gas, both localised and globally (see Figure~\ref{fig:overview}). Observations of representative samples of AGN of different luminosities, hosted in galaxies with diverse properties, are needed to quantify the relevance of the previously mentioned factors on the coupling (see Section~\ref{sec:gasimpact}). 
    
    
    \item Whilst the current AGN state of a galaxy is not a priori a useful proxy for assessing the impact on {\em global} galaxy properties, by studying currently active AGN, with spatially-resolved observations, we can obtain crucial information on the physics of {\em localised} impact. This is essential to determine how the energy couples with the gas, and under which circumstances it enhances or reduces star formation efficiency (see Figure \ref{fig:overview} and Section~\ref{sec:localimpact}).
    
    \item Evidence of the cumulative impact of AGN episodes on global galaxy properties is likely found in the galaxy population as a whole (not necessarily currently active). Distributions of galaxy properties are important for testing, and ruling out, different AGN feedback prescriptions implemented in cosmological simulations (see Section~\ref{sec:longterm}). 
    
\end{itemize}

Over the next decade we expect an explosion of data that will build up an increasingly complete observational picture of how AGN couple to multi-phase gas, and the resulting impact of this interaction. With the latest observatories and instrumentation, this will be possible for ever-more representative galaxy samples, including for the relatively unexplored lowest mass galaxies \cite{davis20,bohn21,koundami22,arjonaGalvez24} and those in the earliest Universe \cite{carniani23,tripodi24}. High-resolution simulations, that incorporate more physics, and that can be compared meaningfully to observations (e.g., \cite{costa22,tanner22,meenakshi22,audibert23,bourne23,mercedesfeliz23,dutta24}) provide a promising future for understanding physics of the {\em localised and transient} processes of interactions between AGN and their host galaxies, and the potential {\em cumulative, global, and long-term impact} of these interactions. All aspects of the multi-faceted observational approaches, summarised in Figure~\ref{fig:overview}, remain important as we strive for a comprehensive understanding of the role of AGN feedback in galaxy evolution.

\vspace{5pt} 

\authorcontributions{The conceptualization, research and writing were carried out jointly by C.M.H and C.R.A.}

\funding{C.M.H acknowledges funding from a United Kingdom Research and Innovation grant (code: MR/V022830/1). C.R.A thanks support by the EU H2020-MSCA-ITN-2019 Project 860744 ``BiD4BESt: Big Data applications for black hole Evolution STudies'' and from project PID2022-141105NB-I00 ``Tracking active galactic nuclei feedback from parsec to kiloparsec scales'', funded by
MICINN-AEI/10.13039/501100011033.}

\dataavailability{``Not applicable''} 

\acknowledgments{We thank the participants of ``AGN on the Beach'' and ``The Importance of Jet-induced Feedback on Galaxy Scales'', meetings for inspiring the contents of this article. We thank Ilaria Ruffa for the organisation of the first meeting and both the organising committee and the Lorentz Center staff for the latter. We thank Houda Haidar and Tiago Costa for providing insightful comments and suggestions and Gabriel Pérez Díaz (IAC MultiMedia Service, SMM) for producing the digital version of Figures 1, 2 and 3. We thank the referees for their reports which helped us improve the clarity of this article.
}

\conflictsofinterest{The authors declare no conflict of interest. The funders had no role in study or in the decision to publish the~results.}






\begin{adjustwidth}{-\extralength}{0cm}

\reftitle{References}

\end{adjustwidth}
\end{document}